\documentclass[journal]{IEEEtran}

\usepackage[cmex10]{amsmath}
\usepackage{cite}
\usepackage{amssymb}
\usepackage{graphicx}

\newtheorem{teorema}{Theorem}
\newtheorem{proposicao}{Proposition}

\newtheorem{lema}{Lemma}
\newtheorem{definicao}{Definition}

\newtheorem{exemplo}{Example}

\newcommand{\F}{\mathbb{F}}
\newcommand{\N}{\mathbb{N}}
\newcommand{\C}{\mathcal{C}}
\newcommand{\PO}{(P,\pi)}
\newcommand{\PP}{(\overline{P},\pi)}

\begin{document}

    \title{Classification of poset-block spaces admitting MacWilliams-type identity}

\author{Jerry~Anderson~Pinheiro~and~Marcelo~Firer
\thanks{J. A. Pinheiro is with IMECC--UNICAMP, State University of Campinas, CEP 13083-859, Campinas,
SP, Brazil (e-mail: jerryapinheiro@gmail.com).}
\thanks{M. Firer is with IMECC--UNICAMP, State University of Campinas, CEP 13083-859, Campinas,
SP, Brazil (e-mail: mfirer@ime.unicamp.br).}
}


\maketitle

    \begin{abstract}
        In this work we prove that a poset-block space admits a MacWilliams-type identity if and only if the poset is hierarchical 
and at any level of the poset, all the blocks have the same dimension. When the poset-block admits the 
MacWilliams-type identity we explicit the relation between the weight enumerators of a code and its dual.
    \end{abstract}


\section{Introduction}
Due to both the interest in generalizing classic problems in coding theory and to applications in cryptography, experimental 
designs and high-dimensional numerical integration (see for example~\cite{comb} and~\cite{bloco}), by the mid 1990s 
researches began to study codes considering metrics others than the usual Hamming metric over $\F_q^n$. Among those 
families of metrics are the poset metrics~\cite{brualdi} and the block metrics~\cite{bloco}. Much of 
the classical theory has been generalized to codes in spaces endowed with a poset metric, as can be seen, for example, 
in~\cite{charac},~\cite{kim},~\cite{cons} and~\cite{groups}. 

In 2008 Firer \textit{et al}~\cite{m1} presented the family of metrics called poset-block that generalizes all the previous ones. 
In this work we generalize to poset-block spaces the 
characterization given in~\cite{kim} for poset-metric spaces of poset-block metrics admitting MacWilliams-type identity.

Let $[m]:=\{1,2,\cdots ,m\}$ be a finite set. If $\preccurlyeq$ is a partial order relation in $[m]$, we say 
$P:=([m],\preccurlyeq)$ is a \textit{poset} and denote by $\preccurlyeq_{P}$ the order in $P$. An \textit{ideal} in a poset is a 
nonempty subset  $I\subset [m]$ such that, for $i\in I$ and $j\in [m]$, if $j\preccurlyeq_P i$ then $j\in I$. Given 
$A\subset [m]$, we denote by $\langle A\rangle_P$ the smaller ideal of $P$ containing $A$. If $A=\{i\}$, we will denote 
by $\langle i\rangle_P$ the ideal $\langle\{i\}\rangle_P$. A \textit{chain} in a poset $P$ is a subset of $[m]$ such that 
every two elements are comparable.

Let $\F_q$ be a finite field and $\F_q^n$ the vector space of $n$-tuples over $\F_q$. Given $m\in[n]$, $P$ a poset over 
$[m]$ and $\pi :[m]\rightarrow \N$ a map such that $n=\sum_{i=1}^m \pi(i)$,  we say that $\pi$ is a \textit{labeling} of 
the poset $P$ and that the pair $\PO$ is a \textit{poset-block} structure over $[m]$.

We denote $k_i=\pi(i)$, and consider the vector space over $\F_q$
$$V:= \F_q^{k_1}\times\F_q^{k_2}\times\cdots \times\F_q^{k_m},$$
isomorphic to $\F_q^n$. Given $u\in \F_q^n$, there is a unique decomposition $u=(u_1,\cdots ,u_m)$ with $u_i\in \F_q^{k_i}$, 
$i\in [m]$. The $\pi$\textit{-support} and the $\PO$\textit{-weight} of $u$ are defined respectively as
$$supp_{\pi}(u):=\{i\in[m]:u_i\neq 0\in\F_q^{k_i}\}$$
and
$$w_{\PO}(u):=|\langle supp_{\pi}(u)\rangle_P|,$$
where $|.|$ denotes the cardinality of the given set. For $u,v\in \F_q^n$,
$$d_{\PO}(u,v):=w_{\PO}(u-v)$$
defines a metric over $\F_q^n$ called \textit{poset-block metric}, or just $\PO$\textit{-distance} between $u$ and $v$.

We note that when $\pi(i)=1$ for every $i\in[m]$ the $\PO$-distance is usual poset distance introduced in \cite{brualdi}, while 
imposing $P$ to be a trivial poset ($i\preccurlyeq j\iff i=j$) turns the $\PO$-distance into the block distance defined in 
\cite{bloco}. Interweaving the poset and the block structures opens a wide range of possibilities for searching for codes with 
interesting metric characteristics, such as perfect codes, since poset and block metrics have opposite effects on 
distances: while enlarging the relations on a poset enlarges the distances (hence ``shrinks'' metric balls), enlarging the blocks 
diminishes distances (hence ``blows'' metric balls).

Concerned with MacWilliams-type identities, dual posets play a crucial role: 

\begin{definicao}
Given a poset $P$ over $[m]$, the \textit{dual poset} is the poset $\overline{P}$ defined by the relations 
$$i\preccurlyeq_{P}j\iff j\preccurlyeq_{\overline{P}}i$$
for every $i,j\in [m]$. The pair $\PP$ is called the \textit{dual poset-block}. 
\end{definicao}

Given $j\in [m]$, the \textit{rank} of $j$, denoted by $h_P(j)$, is 
$$h_P(j):=max\{|C| : C \subset \langle j\rangle_P\ \mbox{and} \ C\ \mbox{is a chain}\}.$$ 
The \textit{height} $h(P)$ \textit{of} $P$ is the maximal rank of the elements of $[m]$. The $i$\textit{-level} of $P$ is 
$\Gamma_P^i:=\{j\in [m]: h_P(j)=i\}$. We define $b_i=\sum_{j\in \Gamma_P^i} k_j$ as the sum of the dimensions of the blocks 
associated by $\pi$ to the $i$-level  of $P$, and we call it the \textit{dimension of} $\Gamma_P^i$.

A poset-block $\PO$ is said to be \textit{hierarchical} if given $j_1\in \Gamma_P^i$ we have that $j_1 \preccurlyeq_{P} j$ 
for all $j\in \Gamma_P^{i+1}$. Defining a hierarchical poset on $[m]$ is equivalent to choosing an ordered partition of 
$[m]$ (the partition defined by the different levels), thus it is a quite large set of posets (or poset metrics) including, 
as a particular case, the block structures presented in \cite{bloco} when the poset structure is trivial 
($h(P)=1$), the Niederreiter-Rosenbloom-Tsfasman metric (see \cite{mmetric}) with a unique chain when 
$h(P)=m$ and the block structure is trivial ($k_{i}=1$ for every $i\in[m]$) and the usual 
Hamming structure when both the poset and the block structures are trivial.

Given a poset-block $\PO$ over $[m]$ such that $|\Gamma_P^i|=m_i$, let $\sigma$ be a permutation of $[m]$ such that $\{\sigma^{-1}(r_i+1),\cdots ,\sigma^{-1}(r_i+m_i)\}=\Gamma_P^i$ where $r_i=m_1+\cdots +m_{i-1}$ and $m_0=0$. We let $P_1$ be the poset induced by $\sigma$, ie, the poset in which $\sigma(j_1)\preccurlyeq_{P_1}\sigma(j_2)$ if $j_1\preccurlyeq_P j_2$. Obviously, $P_1$ and $P$ are  isomorphic posets. If we put $\pi_1(i)=\pi(\sigma^{-1}(i))=k_i^\prime$, then the map
$$g:   (\F_q^{k_1}\times\cdots \times \F_q^{k_m}, d_{\PO})   \rightarrow     (\F_q^{k_1^\prime}\times\cdots \times \F_q^{k_m^\prime}, d_{(P_1,\pi_1)})$$
$$ \ \ \ \ \ \ \ \ (v_1,\cdots ,v_m)                                         \mapsto         (v_{\sigma(1)},\cdots ,v_{\sigma(m)})$$
is, by construction, a linear isometry. Hence, up to a linear isometry, we can and will assume that $\Gamma_P^i=\{r_i+1,\cdots ,r_i+m_i\}$, and in this case we say   $\PO$ has a \textit{natural labeling}. Hence, given $u\in \F_q^n$ we may decompose it as
$$u=\sum_{i=1}^{h(P)}\sum_{j=1}^{m_i}\sum_{l=1}^{k_{(r_i+j)}} u_{r_i+j}^le_{s(i,j,l)}$$
where $u_{r_i+j}^l\in \F_q$ are scalars and
$$\left\{ e_{s(i,j,l)}:1\leqslant l \leqslant k_{(r_i+j)}, 1\leqslant j\leqslant m_i,1\leqslant i\leqslant h(P)\right\}$$
is the usual basis of $\F_q^n$, with $s(i,j,l)=l+\sum_{t=0}^{r_i+j-1}k_t$ and $k_0=0$.

A $[n,k,\delta]_q$ linear $\PO$-\textit{code} is a $k$-dimensional subspace $\C \subset \F_q^n$ where $\F_q^n$ is equipped with the poset-block metric $d_{\PO}$ and
$$\delta=min\{w_{\PO}(v): 0\neq v\in \C\}$$
is the $\PO$\textit{-minimum distance} of $\C$. 
\begin{definicao}
Let $\C$ be a linear $\PO$-code.  Its \textit{dual code} is defined as
$$\C^\perp = \{x\in \F_q^n : x\cdot u =0 \ \forall \ u\in \C\}$$
where $x\cdot u$ is the usual formal inner product. We remark that $\C^\perp$ is an $(n-k)$-dimensional linear code. Along this work, $\mathcal{C}^{\perp}$ is considered to be a linear 
$\PP$-code with parameters $[n,n-k]_{q}$ and we denote by $\delta^{\perp}$ its minimal distance (according to the 
$\PP$-metric).
\end{definicao}

Given a linear $\PO$-code $\C$, the $\PO$\textit{-weight enumerator of} $\C$  is the polynomial
$$W_{\C,\PO}(x)=\sum_{u\in\C}x^{w_{\PO}(u)}=\sum_{i=0}^{m}A_{i,\PO}(\C)x^{i},$$
where $A_{i,\PO}(\C)=|\{u\in \C: w_{\PO}(u)=i\}|$. When no confusion may arise, we will use a simplified notation for those coefficients: $A_{i}=A_{i,\PO}(\C)$ and $\overline{A}_i=A_{i,\PP}(\C^{\perp})$. 

Note that
$$V:=\F_q^{b_1}\times \cdots  \times \F_q^{b_t}$$
is a vector space over $\F_q$ isomorphic to $\F_q^n$, so that given $u\in \F_q^n$ we can write $u=(u^1,\cdots ,u^t)$ where $u^i \in \F_q^{b_i}$ and $u^i=(u_{r_i+1},\cdots ,u_{r_i+m_i})$ is such that $u_{r_i+j}\in \F_q^{k_{(r_i+j)}}$.

If $P$ is a poset with $t$ levels, the \textit{leveled} $\PO$\textit{-weight enumerator of} $\C$ is the formal expression
$$W_{\C,\PO}(x;y_0,\cdots ,y_t):=\sum_{u\in \C}x^{w_{\PO}(u)}y_{s_{P}(u)},$$
where $s_{P}(u)=max\{i:  u^i\in\F_q^{b_i}\backslash \{0\}\}$ and $s_{P}(0)=0$. This definition is similar to the one used in~\cite{kim} in the classification of poset metrics that admits MacWilliams-type identity, ie, the case where the block structure is trivial. It is clear that $W_{\C,\PO}(x)=W_{\C,\PO}(x;1,\cdots ,1)$.

\begin{definicao}
    We say that a poset-block $\PO$ admits a MacWilliams-type identity (\textit{MW-I}) if the $\PP$-weight enumerator 
of $\C^\perp$ is uniquely determined by the $\PO$-weight enumerator of $\C$ for every linear $\PO$-code $\C$.
\end{definicao}

MacWilliams-type identities in the context of poset codes have interested researchers (see~\cite{charac},~\cite{gut} 
and~\cite{macty}) since they establish a relation between important invariants of a high information rate code with 
those of a low dimension code, that are much easier to compute. In 2005, Kim and Oh~\cite{kim} proved that a poset space 
admits a MW-I if and only if the poset is hierarchical. In this work we extend this result to the instances that remained open: 
the instance of poset-block (and block metrics as a particular case).

\section{MacWilliams-type identity in $\PO$ spaces}

The example below shows that the condition established in~\cite{kim} is not sufficient to ensure MacWilliams-type identity
in $\PO$ spaces.

\begin{exemplo}\label{example1}
    Let $P=\{1,2,3\}$ be the hierarchical poset with partial order defined by the relations $1\preccurlyeq_P 2$ and
$1\preccurlyeq_P 3$ so that the dual poset  $\overline{P}$ is defined by the relations $2\preccurlyeq_{\overline{P}} 1$
and $3\preccurlyeq_{\overline{P}} 1$. Define $\pi:[3]\rightarrow \N$ by $\pi(1)=1$, $\pi(2)=1$ and $\pi(3)=2$. Then,
direct computations shows that the linear codes
    $$\C_1=\{(0,0,0,0),(0,0,1,0)\}$$
    and
    $$\C_2=\{(0,0,0,0),(0,1,0,0)\}$$
over $\F_2^4$ has the same $\PO$-weight enumerator: $$W_{\C_1,\PO}(x)=1+x^2=W_{\C_2,\PO}(x).$$ However, $$W_{\C_1^{\perp},\PP}(x)=1+2x+x^2+4x^3$$ and $$W_{\C_2^{\perp},\PP}(x)=1+3x+4x^3,$$ so that  MW-I does not hold.
\end{exemplo}

\subsection{Necessary condition for MacWilliams-type identity}
Let $\PO$ be a poset-block in $[m]$ with $t$ levels such that $|\Gamma_P^i|=m_i$ for $i\in [t]$. The three lemmas below are the equivalent, for the poset-block case, of Lemmas (2.1)--(2.4) in~\cite{kim}. Despite the fact their proofs for poset-block being more delicate than in the case of posets (where the blocks are trivial), they are quite similar.
\begin{lema}\label{lema2i}
    Given $u\in \F_q^n$ then $w_{\PP}(u)=m\Leftrightarrow supp_{\pi}(u)\supset \Gamma_P^1$. Furthermore, if $u$ satisfies $supp_{\pi}(u)\subset \Gamma_P^1$, we have that
$$q^{n-b_1}\ \mbox{\large\textbar} \ |\{v\in \F_q^n : u\cdot v=0 \mbox{ and } w_{\PP}(v)=m\}|.$$
where $a\mbox{\large\textbar}b$ means $a$ divides $b$ and $b_1$ is the dimension of $\Gamma_P^1$.
\end{lema}

\begin{IEEEproof}
    The first affirmation is evident. Let $u\in\F_q^n$ such that $supp_{\pi}(u)\subset \Gamma_P^1$. Without loss of generality we 
can assume that $\Gamma_P^1=[m_1]$ and $u=(u_1,\cdots ,u_i,0,\cdots ,0)$ where $i\leqslant m_1$ and $u_j\in \F_q^{k_j}\backslash\{0\}$ for all $j\in [i]$. Set
\begin{align*}
  A:= \{ (v_1,\cdots ,v_i) : v_j\in \F_q^{k_j} \backslash\{0 & \}  \ \forall \ j\in [i] \mbox{ and }\\
                                           & u_1\cdot v_1+\cdots +u_i\cdot v_i=0\}.
\end{align*}
In each $\F_q^{k_j}$ space we have $q^{k_j}-1$ non null vectors, then we have $\prod_{j=i+1}^{m_1}(q^{k_j}-1)$ possibilities of vectors in the blocks associated to elements of the subset $\{i+1,\cdots ,m_1\}$ of $[m]$, since we do not impose restrictions in the $m-m_1$ remaining blocks, by first claim it follows that
\begin{align*}
  |\{v\in \F_q^n : u\cdot v=0 \mbox{ and } w_{\PP}(v) & =m\}|=\\
  & q^{n-b_1}|A|\prod_{j=i+1}^{m_1}(q^{k_j}-1).
\end{align*}
\end{IEEEproof}

\begin{lema}\label{lema3i}
    If a poset-block  $\PO$ admits a  MW-I, then  $j\preceq_{P} i$ for every $i\in \Gamma_P^2$ and $j\in \Gamma_P^1$.
\end{lema}

\begin{IEEEproof}
Assuming $\Gamma_P^2 \neq \emptyset$, it follows that $m>m_1$. Suppose there is $i\in \Gamma_P^2$ that is not comparable 
to some $j\in \Gamma_P^1$, that is, such that $|\langle i\rangle_P|<1+|\Gamma_P^1|$. In this instance there are $u,v\in \F_q^n$ 
such that $supp_{\pi}(u)=\{i\}$, $supp_{\pi}(v)\subset \Gamma_P^1$ and 
$|\langle supp_{\pi}(u)\rangle_P|=|\langle supp_{\pi}(v)\rangle_P|$. Without loss of generality we can admit that 
$u=e_{s(2,1,1)}$. If $\C_u$ and $\C_v$ are two one-dimensional linear $\PO$-codes generated by $u$ and $v$ respectively, then 
$\C_u$ and $\C_v$ have same $\PO$-weight enumerator. Assuming the MW-I in $\PO$, $\C_u^\perp$ and $\C_v^\perp$ must have the same 
$\PP$-weight enumerator. If $x\in \C_u^\perp$ then $x_{r_2+1}^1=0$. Furthermore, by Lemma~\ref{lema2i} $w_{\PP}(x)=m$ if and only 
if $\Gamma_P^1\subset supp_{\pi}(x)$, so that
\begin{align*}
  |\{x\in \C_u^\perp : & w_{\PP}(x)=m\}|= \\
  & \ \ \ \  |\{x\in \F_q^n : x_{r_2+1}^1=0 \mbox{ and } \Gamma_P^1\subset supp_{\pi}(x)\}|.
\end{align*}
Set 
$$A:=\{x_i\in \F_q^{k_i} : x_{r_2+1}^1=0\}$$
and 
$$B:=\{(x_1,\cdots ,x_m) : x_j\neq 0 \ \forall \ j\in[m_1]  \mbox{ and } x_i=0\}.$$
Since $i\notin \Gamma_P^1$, $|A|=q^{k_i-1}$ and $|B|=q^{n-k_i-b_1}\prod_{j=1}^{m_1}(q^{k_j}-1)$, it follows that
\begin{align}
  |\{x\in \C_u^\perp : w_{\PP}(x)=m\}|= & |B||A|=\nonumber \\
  & q^{n-b_1-1}\prod_{j=1}^{m_1}(q^{k_j}-1).\label{e:eqqaution2}
\end{align}
On the other hand
\begin{align}
  \{x\in \C_v^\perp : & w_{\PP}(x)=m\}=\nonumber \\
  & \ \ \{x\in\F_q^n : x\cdot v=0 \mbox{ and } w_{\PP}(x)=m\},\label{e:eqqaution3}
\end{align}
hence, by Lemma~\ref{lema2i} and by Equations (\ref{e:eqqaution2}) and (\ref{e:eqqaution3}) it follows that
$$q\ \mbox{\large\textbar}\prod_{j=1}^{m_1}(q^{k_j}-1),$$
a contradiction because $q$ is power of a prime. Therefore $|\langle i \rangle_P|=1+|\Gamma_P^1|$, ie, $j\preceq_{P} i$ for all $j\in \Gamma_P^1$.
\end{IEEEproof}
Let $P^j= P \backslash \cup_{i=1}^j \Gamma_P^i$. Consider on $P^j$ the order induced by $P$ and let
$\pi^j=\pi|_{[m]\backslash \cup_{i=1}^j \Gamma_P^i}$ be the restriction of $\pi$ to $[m]\backslash \cup_{i=1}^j \Gamma_P^i$.
\begin{lema}\label{lema4i}
    If a poset-block $\PO$ admits the MW-I, then the poset-block $(P^1,\pi^1)$ also admits.
\end{lema}

\begin{IEEEproof}
If $m=m_1$ we have that $[m]\backslash \Gamma_P^1=\emptyset$ and there is nothing to be proved. Let us assume that $m>m_1$ and 
let $\C_1'$ and $\C_2'$ be linear $(P^1,\pi^1)$-codes with length $n-b_1$ and same $(P^1,\pi^1)$-weight enumerator. 
For $i=1,2$, let
$$\C_i:=\F_q^{b_1}\oplus \C_i'= \{(u,v):u\in\F_q^{b_1} \mbox{ and } v\in \C_i'\}$$
be linear $\PO$-codes with length $n$ and same $\PO$-weight enumerator. Since $\PO$ admits MW-I, $\C_1^\perp$ and $\C_2^\perp$ 
have the same $\PP$-weight enumerator. Furthermore, the dual codes $\C_1^\perp$ and $\C_2^\perp$ can be described as
\begin{align*}
  \C_i^\perp=\{(u,v)\in \F_q^{b_1}\times \F_q^{n-b_1}:(u,v) & \cdot(a,b)=0 \\
  & \forall \ a\in \F_q^{b_1} \mbox{ and }b\in \C_i'\}.
\end{align*}
Being $b\in\C_i'$ the null code-word of $\C_i'$, by definition of $\C_i^\perp$ it follows that $u$ is the null element of $\F_q^{b_1}$, hence
$$\C_i^\perp=\{(u,v):u=0 \in \F_q^{b_1}\mbox{ and }v\in \C_i'^{\perp}\}.$$
Therefore, by puncturing the codes $\C_1^\perp$ and $\C_2^\perp$ in the first $b_1$ coordinates, it follows that $ \C_1'^{\perp}$ and 
$\C_2'^{\perp}$ have the same $(P^1,\pi^1)$-weight enumerator.
\end{IEEEproof}

By induction, using Lemmas \ref{lema3i} and \ref{lema4i} we have the following necessary condition for a poset-block $\PO$ to admit a MW-I.

\begin{proposicao}\label{teorema2i}
    If $\PO$ admits the MW-I, then $P$ is a hierarchical poset.
\end{proposicao}

By Example~\ref{example1} we can conclude that the previous condition is not sufficient to assure an MW-I and the following is also necessary:

\begin{proposicao}\label{teorema22i}
    Suppose that $\PO$ admits a MW-I. Then, $\pi(j_1)=\pi(j_2)$ for all $j_1,j_2\in \Gamma_P^i$ and every $1\leqslant i\leqslant h(P)$, ie,  blocks at the same level have the same dimension.
\end{proposicao}

\begin{IEEEproof}
Given $i\in [h(P)]$ consider $j_1,j_2\in\Gamma_P^i$ and assume $\pi(j_1)\leqslant\pi(j_2)$. Let $\C_u$ and $\C_v$ be the one-dimensional linear $\PO$-codes with length $n$ generated by $u=e_{s(i,j_1-r_i,1)}$ and $v=e_{s(i,j_2-r_i,1)}$ respectively, where $r_i=m_1+\cdots+m_{i-1}$. By Proposition \ref{teorema2i} the poset $P$ is hierarchical, and since there are
$$(q^{k_{j_1}-1}-1)+\displaystyle\sum_{\genfrac{}{}{0pt}{}{j\in \Gamma_P^i}{j\neq j_1}}(q^{k_j}-1)$$
elements in $\C_u^\perp$ with support contained in a unique block at the $i$-level of $P$, then
\begin{align*}
    A_{m_{i+1}+\cdots +m_t+1,(\overline{P},\pi)}(\C_u^\perp)=& (q^{k_{j_1}-1}-1)\prod_{\genfrac{}{}{0pt}{}{j\in\Gamma_{P}^l}{i<l\leqslant t}}q^{k_j}\\
    &+\displaystyle\sum_{\genfrac{}{}{0pt}{}{j\in \Gamma_P^i}{j\neq j_1}}(q^{k_j}-1)\prod_{\genfrac{}{}{0pt}{}{j\in\Gamma_{P}^l}{i<l\leqslant t}}q^{k_j}
\end{align*}
since, when considering the dual poset $\overline{P}$, there are no restrictions on the coordinates in the blocks belonging to levels higher (in $P$) than $i$. In a similar way we find that
\begin{align*}
    A_{m_{i+1}+\cdots +m_t+1,(\overline{P},\pi)}(\C_v^\perp)=& (q^{k_{j_2}-1}-1)\prod_{\genfrac{}{}{0pt}{}{j\in\Gamma_{P}^l}{i<l\leqslant t}}q^{k_j}\\
    &+\displaystyle\sum_{\genfrac{}{}{0pt}{}{j\in \Gamma_P^i}{j\neq j_2}}(q^{k_j}-1)\prod_{\genfrac{}{}{0pt}{}{j\in\Gamma_{P}^l}{i<l\leqslant t}}q^{k_j}.
\end{align*}
Assuming that $\PO$ admits a MW-I it follows that 
$$A_{m_{i+1}+\cdots +m_t+1,(\overline{P},\pi)}(\C_u^\perp)=A_{m_{i+1}+\cdots +m_t+1,(\overline{P},\pi)}(\C_v^\perp),$$
ie, $\pi(j_1)=\pi(j_2)$.
\end{IEEEproof}

From the two previous propositions it follows that:

\begin{teorema}\label{t:terr}
    If $\PO$ admits a MacWilliams-type identity then $P$ is a hierarchical poset and blocks at the same level have the same dimension.
\end{teorema}

\begin{figure}[htbp]
	\centering
		\includegraphics[width=0.2\textwidth]{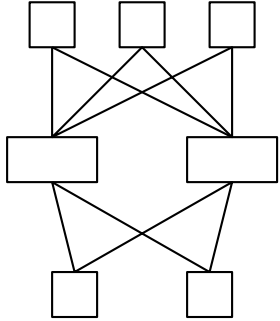}
	\caption{Diagram of a typical hierarchical poset-block with blocks of equal dimension at each level.}
	\label{fig:figure2}
\end{figure} 

\subsection{Sufficient condition for MacWilliams-type identity}
In this section we will prove that the conditions found to be necessary will also be sufficient.
Let $\PO$ be a hierarchical poset-block over $[m]$ with $t$ levels such that $|\Gamma_P^i|=m_i$, with $i\in [t]$. As before, we let $m_0=0$ and $r_i=m_1+\cdots +m_{i-1}$. We can assume without loss of generality that $\Gamma_P^i=\{r_i+1,\cdots ,r_i+m_{i}\}$. Let $d_i=\pi(r_i+j)$ for every $j\in[m_i]$, ie, blocks at the same level have the same dimension. Under this condition the dimension of the $i$-level is given by
$$b_i=\sum_{j=1}^{m_i}\pi(r_i+j)=m_id_i.$$

We note that $n=b_1+\cdots +b_t$ and $m=m_1+\cdots +m_t$. Given $i\in \{0,1,\cdots ,t\}$, set

\begin{itemize}
	\item $\widehat{b_i}=n-(b_{1}+\cdots +b_i)$;
	\item $\widehat{m_i}=m-(m_{1}+\cdots +m_i)$ and
	\item $\widetilde{u^{i+1}}=(u^{i+1},\cdots ,u^{t}) \in \F_q^{\widehat{b_i}}$.
\end{itemize}

With this definitions we have that $w_{\PP}(u)=\widehat{m_i}+w_{\pi_i}(u^i)$ where $w_{\pi_i}(u^i)$ is the $(\Gamma_P^i,\pi|_{\Gamma_P^i})$-weight of $u^i$, the block weight as introduced in~\cite{bloco}. Given a linear $\PO$-code $\C$, the set
\begin{equation*}
	\C_i=\{u\in \C: \widetilde{u^{i+1}}=0\}
\end{equation*}
is a subcode of $\C$ that can be decomposed as $\C_i=\C_i^0\sqcup \C_i^1$ where
\begin{equation*}
	\C_i^0=\{u \in \C_i: u^i= 0\} \ \mbox{ and } \ \C_i^1=\{u \in \C_i: u^i\neq 0\}.
\end{equation*}

Given $i\in [t]$, \textit{the weight enumerator of the} $i$\textit{-level} of $P$ is defined as
\begin{equation}\label{eqnarray1}
    LW_{\C,\PO}^{(i)}(x):= \sum_{j=1}^{m_i} A_{r_i+j}x^{r_i+j}.
\end{equation}
The coefficients of this polynomial represent the weight distribution of code-words such that its support contains elements in the $i$-level and do not contain elements that are above the $i$-level. If we define $LW_{\C,\PO}^{(0)}(x)=A_{0}$, it is clear that
\begin{equation}\label{E:equationn19}
    W_{\C,\PO}(x;y_0,y_1,\cdots ,y_t)=\sum_{i=0}^t LW_{\C,\PO}^{(i)}(x)y_i.
\end{equation}
If for each $i\in [t]$ we have that $y_j=1$ for $j\leqslant i$ and $y_j=0$ for all $j>i$, then the leveled $\PO$-weight enumerator of $\C$ coincides with the $\PO$-weight enumerator of $\C_i$, hence,
\begin{align}\label{eqnnx10}
	W_{\C_i,\PO}(x)-W_{\C_{i-1},\PO}(x)&=LW_{\C,\PO}^{(i)}(x)\nonumber \\
	                                   &=\sum_{u \in \C_i^1} x^{w_{\PO}(u)}.
\end{align}

We introduce now some concepts related to additive characters, that will be used in the proof in a way similar to what was done first by MacWilliams~\cite{macwilliams} in the classical case and later in the poset case (see~\cite{charac},~\cite{kim} and~\cite{macty}).
\begin{definicao}
    An additive character $\chi$ in $\F_q$ is an homomorphism of the additive group $\F_q$ into the multiplicative group of complex numbers with norm $1$. If $\chi\equiv 1$, we say that $\chi$ is the trivial additive character.
\end{definicao}

\begin{lema}\label{23}
	Let $\chi$ be a non trivial additive character of $\F_q$ and $\alpha$ a fix element of $\F_q^j$. Then
	\begin{equation*}
		\sum_{\beta \in \F_q^j} \chi(\alpha \cdot \beta)= \left\{
		\begin{array}{lc}
			q^j,& \mbox{if } \alpha \mbox{ is null}\\
			0, & \mbox{otherwise}\\
		\end{array}
		\right.
	\end{equation*}
\end{lema}

\begin{lema}\label{24}
	Let $\chi$ be a non trivial additive character of $\F_q$. For any linear code $\C \subset \F_q^n$
	\begin{equation*}
		\sum_{v \in \C} \chi(u\cdot v)= \left\{
		\begin{array}{lcl}
			0,& \mbox{if} & u\in \F_q^n\backslash\C^\perp\\
			|\C|, & \mbox{if} & u\in \C^\perp\\
		\end{array}
		\right.
	\end{equation*}
\end{lema}

\begin{definicao}(Hadamard Transform)
    Let $f$ be a complex function defined in $\F_q^n$. The Hadamard transform of $f$ is
    $$\widehat{f}(u)=\sum_{v\in \F_q^n}\chi(u\cdot v)f(v).$$
\end{definicao}
The proof of the following lemma may be found in~\cite{niede}.

\begin{lema}\label{25}
    (Discrete Poisson Summation Formula) Let $\C\subset \F_q^n$ be a linear code and $f$ a complex function defined on $\F_q^n$. Then
    \begin{equation}\label{E:equation20}
        \sum_{v\in \C^\perp}f(v)=\frac{1}{|\C|}\sum_{u\in \C}\widehat{f}(u).
    \end{equation}
\end{lema}

In case both the block and the poset structures are trivial (the Hamming case), the use of the discrete Poisson summation formula to establish the MacWilliams identity is simple: just consider $f(u)=x^{w_H(u)}$ and apply the discrete Poisson summation formula to the Hadamard transform $\widehat{f}(u)=(1+(q-1)x)^{n-w_H(u)}(1-x)^{w_H(u)}$ (as in~\cite{macwilliams}).
If $f(u)=x^{w_{\PP}(u)}z_{s_{\overline{P}}(u)}$, where $s_{\overline{P}}(u)=min\{i: u^i\in \F_q^{b_i}\backslash \{0\}\}$ and $s_{\overline{P}}(0)=t+1$, then
\begin{equation}\label{eq003}
    \sum_{u\in \C^\perp}f(u)=W_{\C^\perp,\PP}(x;z_{t+1},\cdots ,z_1).
\end{equation}
Therefore we will extend this result determining the Hadamard transform of the function $f(u)=x^{w_{\PP}(u)}z_{s_{\overline{P}}(u)}$.

Given $i\in\{0,\cdots ,t\}$, we set 
$$B_i=\{u\in \F_q^n: u^j=0 \ \forall \ 1\leqslant j \leqslant i \text{ and } u^{i+1}\neq 0\}$$ 
and then
\begin{align*}
\widehat{f}(u)&=\sum_{v\in \F_q^n}\chi(u\cdot v)f(v) \\
    & =\sum_{i=0}^t\sum_{v\in B_i}\chi(u\cdot v)f(v) \\
    & =\sum_{i=0}^t\sum_{v\in B_i}\chi(u\cdot v)x^{w_{\PP}(v)}z_{s_{\overline{P}}(v)}.
\end{align*}

Defining $S_i(u)=\sum_{v\in B_i}\chi(u\cdot v)x^{w_{\PP}(v)}z_{s_{\overline{P}}(v)}$, since $B_t=\{0\}$ it follows that
\begin{equation}
    \widehat{f}(u)=z_{t+1}+\sum_{i=1}^{t}S_{i-1}(u).\label{E:equacao1}
\end{equation}

The proof of the sufficiency condition will be done with the aid of four lemmas that allow us to determine $\sum_{u\in \C}\widehat{f}(u)$ as a function of the leveled weight enumerator of $\C$. From  Equation (\ref{E:equacao1}) and assuming that the poset is hierarchical and that blocks at the same level has the same dimension, we get the following four lemmas.

\begin{lema}\label{l:lemma001}
	To $i\in[t]$, denote $\gamma_{i}=(q^{d_{i}}-1)$, then for all $u\in\F_q^n$ we have that
	\begin{equation*}
		S_{i-1}(u)=z_{i}x^{\widehat{m_{i}}}q^{\widehat{b_{i}}}[(1-x)^{w_{\pi_{i}}(u^{i})}(1+\gamma_{i}x)^{m_{i} -w_{\pi_{i}}(u^{i})}-1]
	\end{equation*}
	if $\widetilde{u^{i+1}}$ is a null vector and $S_{i-1}(u)=0$ if $\widetilde{u^{i+1}}$ is not a null vector.
\end{lema}

\begin{IEEEproof}
Since $P$ is a hierarchical poset, if $v\in B_{i-1}$, then $w_{\PP}(v)=\widehat{m_{i}}+w_{\pi_{i}}(v^{i})$, and we denote $v=(v^1,\cdots ,v^{i},\widetilde{v^{i+1}})$. By definition of $S_{i-1}(u)$ and since a character is an additive homomorphism, we have that
\begin{align}\label{eqn004}
     S_{i-1}(u) = z_{i}x^{\widehat{m_{i}}}\sum_{\widetilde{v^{i+1}} \in \F_q^{\widehat{b_{i}}}} & \chi (\widetilde{u^{i+1}} \cdot \widetilde{v^{i+1}})\times \nonumber \\
     & \sum_{v^{i}\in \F_q^{b_{i}}\backslash\{0\}} \chi(u^{i} \cdot v^{i})x^{w_{\pi_{i}}(v^{i})}.
\end{align}
By Lemma~\ref{23}
\begin{equation}\label{eq001}
	\sum_{\widetilde{v^{i+1}} \in \F_q^{\widehat{b_{i}}}} \chi(\widetilde{u^{i+1}} \cdot \widetilde{v^{i+1}})= \left\{
	\begin{array}{lc}
		q^{\widehat{b_{i}}},& \mbox{if }  \widetilde{u^{i+1}} \mbox{ is null}\\
		0, &  \mbox{otherwise}\\
	\end{array}
	\right.
\end{equation}
Being $r_i=m_1+\cdots +m_{i-1}$ and $\chi$ a non trivial additive character, since $v_{r_i+j}\in \F_q^{d_i}$ for every $j\in\{1,\cdots ,m_i\}$ and $w_{\pi_i}(v^i)=\sum_{j=1}^{m_i}\delta(v_{r_i+j})$ where $\delta(u)$ is the Kronecker function (it returns $1$ if $u$ is not null and $0$ otherwise), it follows that 
\begin{align*}
     \sum_{v^{i}\in \F_q^{b_{i}}} \chi(u^{i}\cdot v^{i}) & x^{w_{\pi_{i}}(v^{i})}= \\
     & \ \ \ \ \ \prod_{j=1}^{m_{i}}\sum_{v_{r_i+j}\in \F_q^{d_{i}}} \chi(u_{r_i+j}\cdot v_{r_i+j}) x^{\delta(v_{r_i+j})}.
\end{align*}
Therefore, if $u_{r_i+j}$ is a null vector, then
$$\sum_{v_{r_i+j}\in \F_q^{d_{i}}} \chi(u_{r_i+j} \cdot v_{r_i+j}) x^{\delta(v_{r_i+j})}=1+\gamma_{i}x.$$
If $u_{r_i+j}$ is not a null vector, since $u_{r_i+j}\notin (\F_q^{d_{i}})^\perp$, then by Lemma~\ref{24}
\begin{align*}
	\sum_{v_{r_i+j}\in \F_q^{d_{i}}} \chi( & u_{r_i+j} \cdot v_{r_i+j}) x^{\delta(v_{r_i+j})}= \\
	&1+x \sum_{v_{r_i+j}\in \F_q^{d_{i}}\backslash \{0\}} \chi(u_{r_i+j}\cdot v_{r_i+j})=1-x,
\end{align*}
hence
\begin{align}
	\sum_{v^{i}\in \F_q^{b_{i}}\backslash\{0\}}\chi(u^{i} & \cdot v^{i})x^{w_{\pi_{i}}(v^{i})}= \nonumber \\
	&(1-x)^{w_{\pi_{i}}(u^{i})}(1+\gamma_{i}x)^{m_{i}-w_{\pi_{i}}(u^{i})}-1.\label{eq002}
\end{align}
The result follows from Equations (\ref{eqn004}), (\ref{eq001}) and (\ref{eq002}).
\end{IEEEproof}

\begin{lema}\label{l:lab3}
    Given $i\in[t]$, define
    $$Q_i(x):=\frac{1-x}{1+\gamma_i x},$$
    $$a_i(x):=q^{\widehat{b_{i}}}\left(\frac{1+\gamma_i x}{x}\right)^{m-\widehat{m_{i}}}(1-x)^{\widehat{m_{i-1}}}$$
    and
    $$c_i(x):=x^{\widehat{m_{i}}}q^{\widehat{b_{i}}}\left(\frac{1-x}{Q_{i}(x)}\right)^{m_{i}},$$
    where $\gamma_i=q^{d_i}-1$. Then,
    \begin{align}\label{eqnprinc}
        \sum_{u\in \C}\widehat{f}(u)=& |\C| z_{t+1}\nonumber \\
        &+\left(\frac{x}{1-x}\right)^{m}\sum_{i=1}^{t}a_i(x)z_{i}LW_{\C,\PO}^{(i)}(Q_{i}(x)) \nonumber \\
        &+\sum_{i=1}^{t}z_{i}c_i(x)|\C_{i-1}|-\sum_{i=1}^{t}z_{i}x^{\widehat{m_{i}}}q^{\widehat{b_{i}}}|\C_{i}|.
    \end{align}
\end{lema}

\begin{IEEEproof}
If $u\notin \C_{i}$ then $\widetilde{u^{i+1}}$ is not a null vector and by Lemma~\ref{l:lemma001} we find that
\begin{align}
      \sum_{u\in \C} & S_{i-1}(u)= \nonumber \\
      &  \sum_{u\in \C_{i}}S_{i-1}(u)+\sum_{u\in \C\backslash \C_{i}}S_{i-1}(u)= \nonumber\\
      &  z_{i}x^{\widehat{m_{i}}}q^{\widehat{b_{i}}}\left[\left(\frac{1-x}{Q_{i}(x)}\right)^{m_{i}}\sum_{u\in \C_{i}} Q_{i}(x)^{w_{\pi_{i}}(u^{i})}-|\C_{i}|\right].\label{E:equation03}
\end{align}
If $u\in \C_{i}^1$, then $w_{\PO}(u)=w_{\pi_{i}}(u^{i})+(m-\widehat{m_{i-1}})$, and if $u\in \C_{i}^0$ we have $w_{\pi_{i}}(u^{i})=0$. Since $\C_{i-1}=\C_{i}^0$, then $|\C_{i-1}|=|\C_{i}^0|$ and hence
\begin{align}
            \sum_{u\in \C_{i}}  Q_{i} (x & )^{w_{\pi_{i}}(u^{i})} = \nonumber \\
            & \sum_{u\in \C_{i}^1}Q_{i}(x)^{w_{\pi_{i}}(u^{i})}+\sum_{u\in \C_{i}^0}Q_{i}(x)^{w_{\pi_{i}}(u^{i})}= \nonumber \\
			& \frac{1}{Q_{i}(x)^{m-\widehat{m_{i-1}}}}\sum_{u\in \C_{i}^1} Q_{i}(x)^{w_{\PO}(u)}+|\C_{i-1}|.
\label{E:equation04}
\end{align}
Since $m-\widehat{m_{i+1}}+m_i=m-\widehat{m_i}$ and by Equation (\ref{eqnarray1}) we have that $\sum_{u\in \C_{i}^1} Q_{i}(x)^{w_{\PO}(u)}=LW_{\C,\PO}^{(i)}(Q_{i}(x))$, by replacing Equation (\ref{E:equation04}) into (\ref{E:equation03}) it follows that
\begin{align}
	\sum_{u\in \C}S_{i-1}(u)& =\left(\frac{x}{1-x}\right)^{m}q^{\widehat{b_{i}}}\left(\frac{1+\gamma_i x}{x}\right)^{m-\widehat{m_{i}}}\times \nonumber \\
	& (1-x)^{\widehat{m_{i-1}}}z_{i}LW_{\C,\PO}^{(i)}(Q_{i}(x)) \nonumber+ \\
                     & +z_{i}x^{\widehat{m_{i}}}q^{\widehat{b_{i}}}\left[\left(\frac{1-x}{Q_{i}(x)}\right)^{m_{i}}|\C_{i-1}| -|\C_{i}|\right]. \label{E:equation07}
\end{align}
By Identity (\ref{E:equacao1}), $\widehat{f}(u)=z_{t+1}+\sum_{i=1}^{t}S_{i-1}(u)$, then by Equation (\ref{E:equation07})
\begin{align*}
                            \sum_{u\in \C}  &\widehat{f}(u )  =  |\C| z_{t+1}+\sum_{i=1}^{t}\sum_{u\in \C}S_{i-1}(u)  \\
                             & =|\C| z_{t+1}+\left(\frac{x}{1-x}\right)^{m}\sum_{i=1}^{t} a_i(x)z_{i} LW_{\C,\PO}^{(i)}(Q_{i}(x)) \\
                              & \ \ \ +\sum_{i=1}^{t}z_{i} c_i(x)|\C_{i-1}| -\sum_{i=1}^{t}z_{i}x^{\widehat{m_{i}}}q^{\widehat{b_{i}}}|\C_{i}|.
\end{align*}
\end{IEEEproof}
In the definition of $W_{\C,\PO}(x;y_0,\cdots ,y_t)$, the $y_i's$ were considered as formal symbols. In two next lemmas we consider specific situations that will determine the weight enumerator in the stated conditions.
\begin{lema}\label{l:lab2}
    Let
    \begin{equation*}
        g_j = \left\{
        \begin{array}{ll}
            \sum_{i=j+1}^{t}c_i(x)z_{i},   & \mbox{if } 0\leqslant j \leqslant t-1\\
            0,  & \mbox{if } j=t
        \end{array}.
        \right.
    \end{equation*}
    Then
    \begin{equation*}
        \sum_{i=1}^{t}z_{i}c_i(x)|\C_{i-1}| = W_{\C,\PO}(1;g_0,\cdots ,g_t).
    \end{equation*}
\end{lema}

\begin{IEEEproof}
Since $r_i=m_1+\cdots +m_{i-1}$ and
\begin{equation}\label{eqtf}
|\C_i| = A_{0}+A_{1}+\cdots +A_{r_i+m_i}=\sum_{j=0}^{i}LW_{\C,\PO}^{(j)}(1),
\end{equation}
then
\begin{align*}
\sum_{i=1}^{t} z_{i}  c_i( & x)  |\C_{i-1}| =\\
                    &A_{0}(c_1(x)z_1+c_1(x)z_2+\cdots +c_{t}(x)z_t) \\
                                      &+(A_{1}+\cdots +A_{m_1})(c_2(x)z_2+\cdots +c_{t}(x)z_t) \\
                                      &+\cdots +\\
                                      &+(A_{m_1+\cdots +m_{t-2}+1}+\cdots +A_{m_1+\cdots +m_{t-1}})c_{t}(x)z_t \\
                                      = &\sum_{i=0}^t LW_{\C, \PO}^{(i)}(1)g_i
\end{align*}
hence the result follows from Identity (\ref{E:equationn19}).
\end{IEEEproof}

The proof of the next lemma is omitted since it follows the same steps as in the proof of Lemma~\ref{l:lab2}.

\begin{lema}\label{l:lab1}
    Let
    \begin{equation*}
        h_j = \left\{
        \begin{array}{ll}
            \sum_{i=j}^{t}z_{i}x^{\widehat{m_{i}}}q^{\widehat{b_{i}}},   & \mbox{if } 1\leqslant j \leqslant t\\
            \sum_{i=1}^{t}z_{i}x^{\widehat{m_{i}}}q^{\widehat{b_{i}}},  & \mbox{if } j=0
        \end{array}.
        \right.
    \end{equation*}
    Then
    \begin{equation*}
        \sum_{i=1}^{t}z_{i}x^{\widehat{m_{i}}}q^{\widehat{b_{i}}}|\C_{i}|=W_{\C,\PO}(1;h_0,\cdots ,h_t).
    \end{equation*}
\end{lema}

Before we proceed to prove the next theorem we recall we are assuming the following collection of conditions and notations:
\begin{itemize}
	\item $\PO$ a poset-block over $[m]$ with $t$ levels;
	\item $P$ is hierarchical;
	\item $r_i=m_1+\cdots +m_{i-1}$;
	\item $\Gamma_P^i=\{r_i+1,\cdots ,r_i+m_i\}$;
	\item $d_i=\pi(r_i+j)$ for every $j\in\{1,\cdots ,m_i\}$;
	\item $b_i=m_id_i$ is such that $\sum_{i=1}^t b_i=n$.
\end{itemize}

Now we can prove that necessary conditions stated in Theorem \ref{t:terr} are also sufficient to have a MW-I.
\begin{teorema}\label{t:terema00001}
    Under the conditions above stated, the poset-block $\PO$ admits a MacWilliams-type identity.
\end{teorema}
\begin{IEEEproof}
By (\ref{E:equation20}) and (\ref{eq003}) we have that
\begin{equation}\label{eqates}
W_{\C^\perp,(\overline{P},\pi)}(x;z_{t+1},\cdots ,z_1)= \frac{1}{|\C|}\sum_{u\in \C} \widehat{f}(u).
\end{equation}
Considering Equation (\ref{E:equationn19}) we have that
$$a_i(x)z_iLW_{\C,(P,\pi)}^{(i)}(Q_i(x))=W_{\C,(P,\pi)}(Q_i(x);y_0,\cdots ,y_t),$$
for every $i\in \{1,\cdots ,t\}$, where $a_i(x)z_i=y_{i}$ and $y_j=0$ for every $j\neq i$. Substituting the identities obtained in Lemma \ref{l:lab2} and Lemma \ref{l:lab1} into Equation (\ref{eqnprinc}) it follows that
\begin{align}
|\C| &W_{\C^\perp,(\overline{P},\pi)}(x;z_{t+1},\cdots ,z_1)=|\C|z_{t+1}\nonumber\\
& \ \ \ +\left(\frac{x}{1-x}\right)^{m} W_{\C,(P,\pi)}(Q_1(x);0,a_1(x)z_1,0,\cdots ,0)\nonumber \\
& \ \ \ +\left(\frac{x}{1-x}\right)^{m} W_{\C,(P,\pi)}(Q_2(x);0,0,a_2(x)z_2,0,\cdots ,0)\nonumber \\
& \ \ \ +\cdots +\left(\frac{x}{1-x}\right)^{m} W_{\C,(P,\pi)}(Q_t(x);0,\cdots ,0,a_{t}(x)z_t) \nonumber \\
& \ \ \ +  W_{\C,(P,\pi)}(1;g_0,\cdots ,g_t)- W_{\C,(P,\pi)}(1;h_0,\cdots ,h_t). \nonumber
\end{align}
On the left side of the above equality we have the leveled weight enumerator of $\C^\perp$ (the dual code of $\C$). On the right side we have an expression that depends not on the code itself but only on the leveled weight enumerator of  $\C$. Hence, if $\C_1$ is a linear $(P,\pi)$-code that has the same $(P,\pi)$-polynomial as $\C$, since $W_{\C_1^\perp,(\overline{P},\pi)}(x;1,\cdots ,1)$ is the $(\overline{P},\pi)$-polynomial of $\C_1^\perp$, it follows that
$$W_{\C_1^\perp,(\overline{P},\pi)}(x;1,\cdots ,1)=W_{\C^\perp,(\overline{P},\pi)}(x;1,\cdots ,1),$$
ie, the $(\overline{P},\pi)$-polynomial of $\C^\perp$ is uniquely determined by $(P,\pi)$-polynomial of $\C$ for every code $\C$, hence the poset-block structure admits a MW-I.
\end{IEEEproof} 

\subsection{Relationship between Weight Distributions}
In this section, we will use the same conditions and notations stated before Theorem~\ref{t:terema00001} in the previous section. For every $k\in\{0,\cdots ,n\}$, let
$$P_k^{\gamma_{i}}(x:n)=\sum_{l=0}^k (-1)^{l}\gamma_{i}^{k-l}\genfrac{(}{)}{0pt}{}{x}{l}\genfrac{(}{)}{0pt}{}{n-x}{k-l}$$
be the Krawtchouk polynomial whose generator function is given by
\begin{equation}\label{funger}
(1+\gamma_{i} z)^{n-x}(1-z)^x=\sum_{k=0}^\infty P_k^{\gamma_{i}}(x:n)z^k.
\end{equation}
If $x\in \{0,\cdots ,n\}$, we can switch the upper limit of summation by $n$. This generator functions arise naturally when we are setting a relationship between the $\PO$-polynomial coefficients of $\C$ and the $\PP$-polynomial coefficients of $\C^\perp$ (for details about the Krawtchouk polynomials in coding theory, see~\cite{macwilliams}).
\begin{lema}\label{lema1j}
Let $\PO$ be a poset-block over $[m]$ that admits MW-I and $\C$ a linear $\PO$-code with length $n$. Then
\begin{align}
| & \C| W_{\C^\perp,\PP}(x)=|\C|+ \nonumber \\
& \sum_{i=1}^{t} q^{\widehat{b_{i}}}x^{\widehat{m_{i}}} \left[\sum_{k=1}^{m_{i}} \left(a_k(j:m_{i}) +\binom{m_{i}}{k}\gamma_{i}^k |\C_{i-1}| \right) x^k \right]\label{e:equation31}
\end{align}
where $a_k(j:m_{i})=\sum_{j=1}^{m_{i}}A_{r_i+j} P_k^{\gamma_{i}}(j:m_{i})$.
\end{lema}

\begin{IEEEproof}
Set
\begin{align*}
E_1 & (x)= \\
& \sum_{i=1}^{t} q^{\widehat{b_{i}}}\left(\frac{1+\gamma_{i}x}{x}\right)^{m-\widehat{m_{i}}}(1-x)^{\widehat{m_{i-1}}}LW_{\C,\PO}^{(i)}(Q_{i}(x))
\end{align*}
and
$$E_2(x)=\sum_{i=1}^{t} x^{\widehat{m_{i}}}q^{\widehat{b_{i}}}\left(\left(\frac{1-x}{Q_{i}(x)}\right)^{m_{i}}|\C_{i-1}|-|\C_{i}|\right).$$
Putting $z_1=\cdots =z_{t+1}=1$ and replacing (\ref{eqnprinc}) in (\ref{eqates}), it follows that
\begin{equation}
W_{\C^\perp,\PP}(x)= \ 1 + F_1(x)+\frac{1}{|\C|} E_2(x)\label{e:equation33}
\end{equation}
where $F_1(x)=\frac{1}{|\C|}\frac{x^m}{(1-x)^m} E_1(x)$. Using the Identity (\ref{eqnarray1}) in $E_1(x)$ and recalling that $r_i=m-\widehat{m_{i-1}}$ and $\widehat{m_i}-\widehat{m_{i-1}}=m_i$, it follows that
\begin{align}
 E_{1}(x)&= \sum_{i=1}^{t} q^{\widehat{b_{i}}}\left(\frac{1+\gamma_{i}x}{x}\right)^{m-\widehat{m_{i}}}(1-x)^{\widehat{m_{i-1}}}\times \nonumber \\
         & \ \ \ \ \ \ \ \ \sum_{j=1}^{m_{i}}A_{r_i+j}\left(\frac{1-x}{1+\gamma_{i}x}\right)^{r_i+j} \nonumber \\
         & = \sum_{i=1}^{t} \frac{q^{\widehat{b_{i}}}}{x^{m-\widehat{m_{i}}}} \sum_{j=1}^{m_{i}}A_{r_i+j}(1+\gamma_{i}x)^{m_{i}-j}(1-x)^{m+j},\nonumber 
\end{align}
and therefore
\begin{align}
F_1(x)&= \frac{1}{|\C|}\sum_{i=1}^{t} q^{\widehat{b_{i}}}x^{\widehat{m_{i}}} \sum_{j=1}^{m_{i}}A_{r_i+j}(1+\gamma_{i}x)^{m_{i}-j}(1-x)^{j}\nonumber \\
        &=\frac{1}{|\C|}\sum_{i=1}^{t} q^{\widehat{b_{i}}}x^{\widehat{m_{i}}} \sum_{j=1}^{m_{i}}A_{r_i+j}\sum_{k=0}^{m_{i}}P_k^{\gamma_{i}}(j:m_{i})x^k
\end{align}
where the second equality follows from (\ref{funger}). Hence if 
$$a_k(j:m_{i})=\sum_{j=1}^{m_{i}}A_{r_i+j} P_k^{\gamma_{i}}(j:m_{i}),$$ 
since $P_0^{\gamma_{i}}=1$, and then $a_0(j:m_{i})=|\C_{i}|-|\C_{i-1}|$ by (\ref{eqtf}). Therefore
\begin{align}
|\C| & F_1(x) =\nonumber \\
 &\sum_{i=1}^{t} q^{\widehat{b_{i}}}x^{\widehat{m_{i}}} \sum_{k=0}^{m_{i}} \left(\sum_{j=1}^{m_{i}}A_{r_i+j} P_k^{\gamma_{i}}(j:m_{i})\right)x^k\nonumber \\
&=\sum_{i=1}^{t} q^{\widehat{b_{i}}}x^{\widehat{m_{i}}}\left( |\C_{i}|-|\C_{i-1}| +\sum_{k=1}^{m_{i}} a_k(j:m_{i}) x^k\right).\label{e:equation35}
\end{align}
From Newton's binomial theorem we have that
\begin{align}
E_2(x)=&\sum_{i=1}^{t}x^{\widehat{m_{i}}}q^{\widehat{b_{i}}}\left[\left(1+\sum_{k=1}^{m_{i}}\binom{m_{i}}{k}\gamma_{i}^k x^k\right)|\C_{i-1}|-|\C_{i}| \right] \nonumber \\
 =&\sum_{i=1}^{t}x^{\widehat{m_{i}}}q^{\widehat{b_{i}}}\left(|\C_{i-1}|-|\C_{i}| +\sum_{k=1}^{m_{i}}\binom{m_{i}}{k}\gamma_{i}^k |\C_{i-1}| x^k\right) \label{e:equation34}
\end{align}
and the result follows from (\ref{e:equation33}), (\ref{e:equation35}) and (\ref{e:equation34}).
\end{IEEEproof}
In the conditions stated in Lemma (\ref{lema1j}) we have that
\begin{align}
                        W & _{\C^\perp, \PP}(x) = \nonumber \\
                          & \overline{A}_{0} + (\overline{A}_{1} x+\cdots +\overline{A}_{m_t} x^{m_t})\nonumber \\
                          &+(\overline{A}_{m_t+1} x+\cdots +\overline{A}_{m_t+m_{t-1}} x^{m_{t-1}})x^{m_t}\nonumber \\
                          &+\cdots +\nonumber \\
                          &+(\overline{A}_{m_t+\cdots +m_{2}+1} x+\cdots +\overline{A}_{m_t+\cdots +m_1} x^{m_1})x^{m_t+\cdots +m_{2}}\nonumber \\
                          &=1+\sum_{i=1}^{t}x^{\widehat{m_{i}}}\sum_{k=1}^{m_{i}}\overline{A}_{\widehat{m_{i}}+k} x^{k}\label{e:equation32}
\end{align}
therefore from (\ref{e:equation31}) and (\ref{e:equation32}) follows the next theorem, that characterizes the weight distribution 
of $\C^\perp$ in terms of the distribution of $\C$.

\begin{teorema} Let $\PO$ be a hierarchical poset-block over $[m]$ with $t$ levels satisfying MW-I and $\C$ a linear $\PO$-code with length $n$ over $\F_q$. Being $\gamma_i=(q^{d_i}-1)$ and $b_j$ the dimension of $\Gamma_P^j$, for any given $i\in[t]$ and $k\in[m_i]$ we have that
\begin{align}
\overline{A}_{\widehat{m_{i}}+k}=&\frac{q^{\widehat{b_{i}}}}{|\C|}\sum_{j=1}^{m_{i}}\left(A_{r_i+j}P_k^{\gamma_i}(j:m_{i})\right) \nonumber \\
   &+\frac{q^{\widehat{b_{i}}}}{|\C|} \binom{m_{i}}{k}\gamma_{i}^k\sum_{j=0}^{r_i}A_{j}. \nonumber
\end{align}
\end{teorema}

We remark that when we consider a trivial structure of blocks, $b_j=m_j$ and $d_j=1$ for all $j\in[t]$, then we have the result obtained in Theorem 4.4 from~\cite{kim}.
On the other hand, when considering a trivial poset structure (an antichain poset where none of elements are comparable), then $t=1$ and $m=m_1$, hence given $k\in [m_1]$ we have that
$$\overline{A}_k=\frac{1}{|\C|}\sum_{j=0}^m A_j P_k^{\gamma_1}(j:m).$$


    \section*{Acknowledgment}
The first author is currently pursuing the M.Sc. degree in the Institute of Mathematics, Statistics and Scientific 
Computing--UNICAMP. He was supported by CAPES. The second author was partially supported by FAPESP, grant $2007$/$56052$--$8$. 
A partial and initial version of this work was will appear in the Proceedings of ITW 2011.

    \bibliographystyle{IEEEtran}
    \bibliography{biliog}

\begin{thebibliography}{10}
\providecommand{\url}[1]{#1}
\csname url@samestyle\endcsname
\providecommand{\newblock}{\relax}
\providecommand{\bibinfo}[2]{#2}
\providecommand{\BIBentrySTDinterwordspacing}{\spaceskip=0pt\relax}
\providecommand{\BIBentryALTinterwordstretchfactor}{4}
\providecommand{\BIBentryALTinterwordspacing}{\spaceskip=\fontdimen2\font plus
\BIBentryALTinterwordstretchfactor\fontdimen3\font minus
  \fontdimen4\font\relax}
\providecommand{\BIBforeignlanguage}[2]{{%
\expandafter\ifx\csname l@#1\endcsname\relax
\typeout{** WARNING: IEEEtran.bst: No hyphenation pattern has been}%
\typeout{** loaded for the language `#1'. Using the pattern for}%
\typeout{** the default language instead.}%
\else
\language=\csname l@#1\endcsname
\fi
#2}}
\providecommand{\BIBdecl}{\relax}
\BIBdecl

\bibitem{comb}
H.~Niederreiter, ``A combinatorial problem for vector spaces over finite
  fields,'' \emph{Discrete Mathematics}, vol.~96, pp. 221--228, 1991.

\bibitem{bloco}
K.~Feng, L.~Xu, and F.~J. Hickernell, ``Linear error-block codes,''
  \emph{Finite Fields and Their Applications}, vol.~12, pp. 638--652, 2006.

\bibitem{brualdi}
R.~A. Brualdi, J.~Graves, and K.~Lawrence, ``Codes with a poset metric,''
  \emph{Discrete Mathematics}, vol. 147, pp. 57--72, 1995.

\bibitem{charac}
D.~S. Kim and D.~C. Kim, ``Character sums and {M}ac{W}illiams identities,''
  \emph{Discrete Mathematics}, vol. 287, pp. 155--160, 2004.

\bibitem{kim}
H.~K. Kim and D.~Y. Oh, ``A classification of posets admitting the
  {M}ac{W}illiams identity,'' \emph{IEEE Transactions on Information Theory},
  vol.~51, no.~4, pp. 1424--1431, Apr. 2005.

\bibitem{cons}
S.~Ling and F.~\"{O}zbudak, ``Constructions and bounds on linear error-block
  codes,'' \emph{Des. Codes Cryptogr.}, vol.~45, pp. 297--316, 2007.

\bibitem{groups}
L.~Panek, M.~Firer, H.~K. Kim, and J.~Y. Hyun, ``Groups of linear isometries on
  poset structures,'' \emph{Discrete Mathematics}, vol. 308, pp. 4116--4123,
  2008.

\bibitem{m1}
M.~M.~S. Alves, L.~Paneck, and M.~Firer, ``Error-block codes and poset
  metrics,'' \emph{Advances in Mathematics of Communications}, vol.~2, no.~1,
  pp. 95--111, 2008.

\bibitem{mmetric}
M.~Y. Rosembloom and M.~A. Tsfasman, ``Codes for m-metric,'' \emph{Problems of
  Information Transmission}, vol.~33, no.~1, pp. 45--52, 1997.

\bibitem{gut}
J.~N. Guti\'{e}rrez and H.~Tapia-Recillas, ``A {M}ac{W}illiams identity for
  poset-codes,'' \emph{Congr. Numer}, vol. 133, pp. 63--73, 1998.

\bibitem{macty}
D.~S. Kim and J.~G. Lee, ``A {M}ac{W}illiams-type identity for linear codes on
  weak order,'' \emph{Discrete Mathematics}, vol. 262, pp. 181--194, 2003.

\bibitem{macwilliams}
F.~J. MacWilliams and N.~J. Sloane, \emph{The Theory of Error-Correcting
  Codes}.\hskip 1em plus 0.5em minus 0.4em\relax Amsterdam, The Netherlands:
  North-Holland, 1977.

\bibitem{niede}
R.~Lidl and H.~Niederreiter, \emph{Finite Fields}, 2nd~ed., ser. Encyclopedia
  of Mathematics and its Applications.\hskip 1em plus 0.5em minus 0.4em\relax
  Cambridge, U.K.: Cambridge University Press, 1997, no.~20.

\end{thebibliography}


\begin{IEEEbiographynophoto}{Marcelo Firer}
received the B.Sc. and M.Sc. degrees in 1989 and 1991 respectively, from State University of Campinas, Brazil, and the Ph.D. degree from the Hebrew University of Jerusalem, in 1997, all in Mathematics.
He is currently an Associate Professor of the State University of Campinas. His research interest includes coding theory, action of groups, semigroups and Tits buildings.
\end{IEEEbiographynophoto}

\begin{IEEEbiographynophoto}{Jerry A. Pinheiro}
receives the B.Sc. in 2006 in Computer Science and in 2008 in Mathematics, from Higher Education Center of Foz do Igua\c cu and Western Parana State University respectively.
He is currently pursuing the M.Sc. degree in the Institute of Mathematics, Statistics and Scientific Computing of the State University of Campinas. His current research interest includes poset-block codes and error-correcting codes.
\end{IEEEbiographynophoto}

\end{document}